\documentclass[journal]{IEEEtran}

\usepackage{graphicx}
\usepackage{multirow}
\usepackage{xurl}

\newcommand{\comment}[1]{}

\begin{document}

\title{Spacecube: A fast inverse hyperspectral georectification system}

\author{Thomas~P.~Watson and Eddie~L.~Jacobs,~\IEEEmembership{Life Senior Member,~IEEE}
\thanks{T. Watson and E. Jacobs are with the Department
of Electrical and Computer Engineering, Herff College of Engineering, University of Memphis, 3720 Alumni Ave, Memphis, TN, 38104 USA e-mail: eljacobs@memphis.edu.}
\thanks{}}

\markboth{}%
{Shell \MakeLowercase{\textit{et al.}}: Bare Demo of IEEEtran.cls for IEEE Journals}

\maketitle

\begin{abstract}
Hyperspectral cameras provide numerous advantages in terms of the utility of the data captured. They capture hundreds of data points per sample (pixel) instead of only the few of RGB or multispectral camera systems. Aerial systems sense such data remotely, but the data must be georectified to produce consistent images before analysis. We find the traditional direct georectification method to be slow, and it is prone to artifacts. To address its downsides, we propose Spacecube, a program that implements a complete hyperspectral georectification pipeline, including our own fast inverse georectification technique, using OpenGL graphics programming technologies. Spacecube operates substantially faster than real-time and eliminates pixel coverage artifacts. It facilitates high quality interactive viewing, data exploration, and export of final products. We release Spacecube's source code publicly for the community to use.
\end{abstract}


\IEEEpeerreviewmaketitle

\section{Introduction}
\IEEEPARstart{H}{yperspectral} cameras sense hundreds of wavelength bands, as opposed to the three bands of an RGB camera or single broad band of a monochrome imager. The large amount of wavelength information in the image enables applications such as agricultural health monitoring via vegetation indices~\cite{thenkabail_hyperspectral_2000}, discrimination of defense targets of interest~\cite{chen_sr_target_2018}, and microplastics detection~\cite{faltynkova_hyperspectral_2021} in soil or water.

There are various techniques and methods by which a hyperspectral camera can accomplish this sensing~\cite{guo_miniaturized_2024,noauthor_cubert_2024,noauthor_ci_2024,noauthor_hinalea_2022,yang_ccd_2003}, along with ways a camera using a particular technique may be calibrated and integrated into a complete aerial sensing system~\cite{k_c_lawrence_calibration_2003,geladi_hyperspectral_2004,warren_data_2014,angel_automated_2019}. The system we study in this paper includes a line-scan (i.e. pushbroom) camera mounted on a multirotor small uncrewed aerial system (sUAS).

Hyperspectral image datasets are known as datacubes: whereas monochrome images have two dimensions (NxM), hyperspectral images add a third spectral dimension (NxMxK). These dimensions are known respectively as N=samples, M=lines, and K=bands, though the order can vary. As the name ``cube" suggests, the spectral dimension (K) is often comparable in size to the other two; the system we study captures datacubes of N=900 samples, M=1000 lines, and K=300 bands.

\begin{figure}
\centering
\includegraphics[height=4cm]{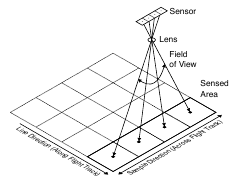}
\caption 
{ \label{fig:geo_diagram}
An aerial line-scan system captures four samples across a line on the ground } 
\end{figure}

\begin{figure}
\centering
\includegraphics[width=3.49in]{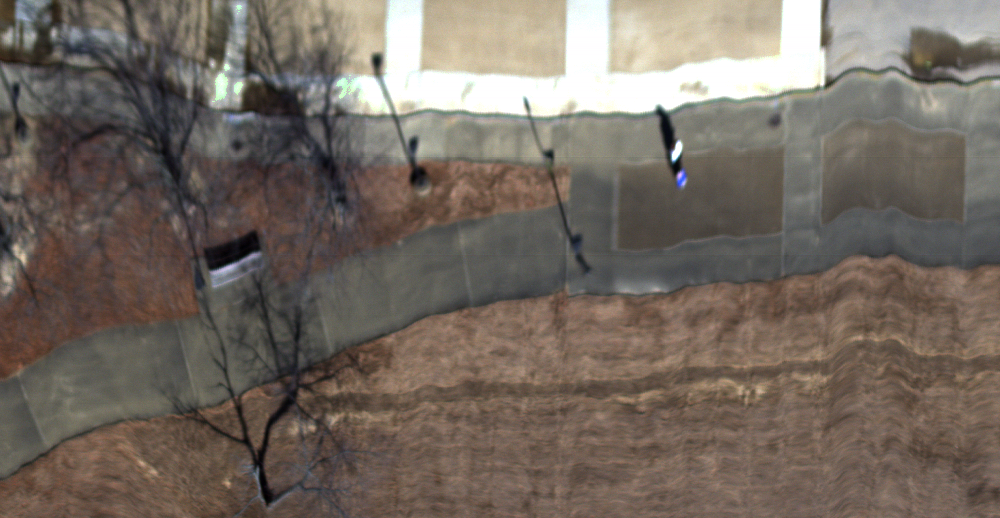}
\caption 
{ \label{fig:geo_before}
Before georectification, with wiggly lines (flight track from left to right) } 
\end{figure}

In an aerial system, the aircraft is programmed to fly along a particular track which covers an area of interest. The line-scan camera is mounted so that the sampled line is perpendicular to the flight track, and programmed to capture lines at regular intervals (Fig.~\ref{fig:geo_diagram}). Simply stacking the lines together has little hope of producing an intelligible 2D image (Fig.~\ref{fig:geo_before}); the motion of the aircraft is quite obvious and causes significant distortion, particularly on multirotor sUAS.

\begin{figure}
\centering
\includegraphics[width=3.49in]{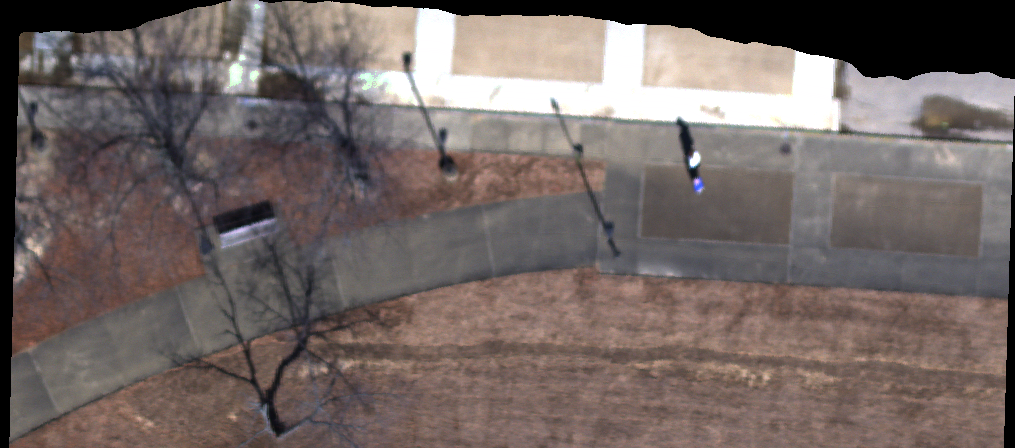}
\caption 
{ \label{fig:geo_after}
After georectification with Spacecube, straight lines are straight (flight track from left to right) } 
\end{figure}

\begin{figure}
\centering
\includegraphics[width=3.49in]{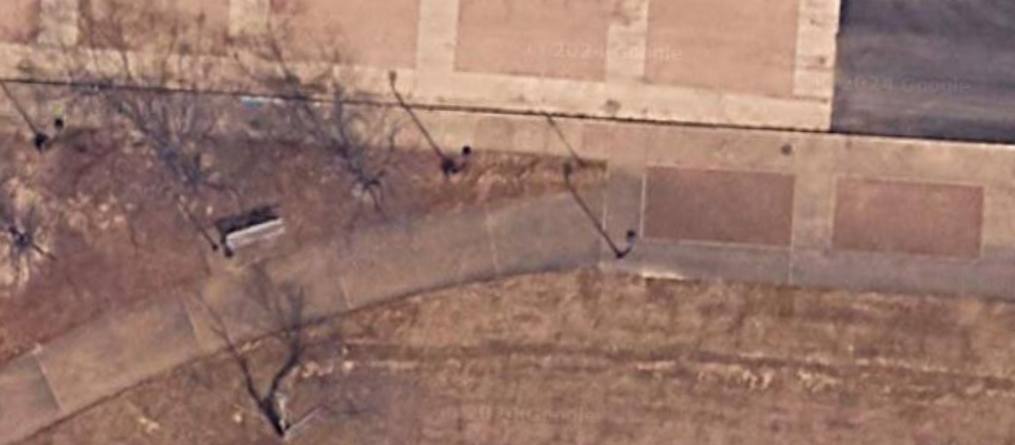}
\caption 
{ \label{fig:geo_google}
Comparable satellite imagery (Imagery/Map Data \copyright 2024 Google) } 
\end{figure}

We therefore need to know the exact location and orientation of the camera when each line was captured. A combination inertial navigation system/global navigation satellite system (INS/GNSS, hereafter just called INS) captures this information. The data is used to compute the geographic position of each sample within each line, then place them correspondingly on an image in a process called georectification~\cite{angel_automated_2019} (Fig.~\ref{fig:geo_after}). In addition to creating local consistency (e.g. showing lines that were straight on the ground as straight in the image), knowing the geographic location of the image and its pixels also enables comparison of different images taken at varying times and by alternate systems, e.g. commercial satellite imagery (Fig.~\ref{fig:geo_google}), even though the exact track will change between flights.

Spacecube is our program that organizes, calibrates, and georectifies hyperspectral line-scan camera and associated INS data. We use OpenGL to implement a mesh-based inverse georectification technique that allows instant preview and rapid export of high quality final datacubes. The technique is used to power an interactive graphical viewer for preview, plus a command line rasterization tool for export. Spacecube's high speed fosters data investigation and exploration, and the mesh rendering eliminates pixel coverage artifacts.

\section{Background\label{sec:background}}

We describe the various ways a hyperspectral camera can be implemented and the reason behind our choice. We also describe the components of a complete hyperspectral system, including calibration and georectification. We then provide a brief explanation of our inverse georectification technique, including how it differs from other techniques. Finally we touch on modern OpenGL graphics processing in preparation for a more thorough explanation of our system in Sec.~\ref{sec:implementation}.

\subsection{Hyperspectral Cameras}
Hyperspectral cameras can use various techniques to accomplish their goal of sensing many wavelengths for each pixel in a scene. This includes micro-filter arrays~\cite{guo_miniaturized_2024} or micro-lens arrays~\cite{noauthor_cubert_2024}, tunable filters~\cite{noauthor_ci_2024,noauthor_hinalea_2022}, and line-scan cameras with diffraction gratings~\cite{yang_ccd_2003}. All discussed cameras are based around an ordinary 2D broadband monochrome image sensor, but additional optics modify the scene before it hits the sensor, and additional processing and/or composition of multiple images is necessary to reconstruct a complete datacube.

Cameras with micro-filter arrays~\cite{guo_miniaturized_2024} place many narrow-band filters in front of the sensor, each of which captures a different part of the spectrum. Depending on the filter structure, each band may not image the same area, so movement might be necessary to form a complete image of the scene in front of the camera. Micro-lens arrays~\cite{noauthor_cubert_2024} function similarly, with each lens also having a filter. With a micro-lens array, all parts of the scene can be captured simultaneously, but sophisticated plenoptic/light field reconstruction is necessary to produce an image. Both types generally have the highest per-frame amount of information, in terms of samples x lines x bands, but are expensive and require complex post-processing.

Cameras with tunable filters simply change the wavelength seen by the entire sensor. This can be done mechanically using a filter wheel~\cite{noauthor_ci_2024}, but is more commonly done using a tunable Fabry-Perot interferometer~\cite{noauthor_hinalea_2022}. This requires time to scan through all the filter configurations and capture a full set of bands. This also assumes the scene does not change and the camera does not move during the scan, making aerial use difficult.

Probably the most basic configuration in terms of capability and cost tradeoff is a line-scan camera, which uses a diffraction grating to split the incoming wavelengths~\cite{yang_ccd_2003}. The grating replaces one spatial axis with the spectral axis. This results in a camera that can capture only one ``line" of the scene at a time, but provides the full spectrum for each sample, producing images of Nx1xM, instead of NxMx1 of the original broadband sensor.

We focus on the line-scan camera as its simple design lends itself to lower cost systems and it is easily applicable to and useful in aerial applications. Spacecube is designed only for this type of hyperspectral camera.

\subsection{Hyperspectral Calibration\label{sec:calibration}}

All types of hyperspectral cameras require calibration to produce data independent of the characteristics of the particular settings, sensor, and scene. The sensor outputs digital numbers (DN) which are proportional to illumination, but there are several effects that must be considered and mitigated to produce an objective output and provide scientific meaning to the data~\cite{k_c_lawrence_calibration_2003,geladi_hyperspectral_2004}.

The sensor can be operated at a variety of framerates, exposure times, and electronic gain values, which all influence the sensor's response (i.e. DN value per unit illumination). These settings are adjusted before and during flight to optimize the capture range and avoid over/under-exposure. They are recorded with the datacube to determine the calibration data to use and correctly scale the final output.

The sensor generates non-zero output values even when no light illuminates it; a calibration dark line taken when the sensor is completely covered is subtracted from each line to mitigate this effect. A line taken of a spatially uniform calibration source with known intensity in each wavelength is then used to convert the value from the sensor to a radiometric quantity (e.g. microflicks). These corrections remove the effects of the sensor's wavelength-dependent response (due to e.g. spectrally varying quantum efficiency) and spatial non-uniformity (due to e.g. lens vignetting).

Spacecube handles loading and applying calibration data and allows the user to interactively select which are applied to explore the effect of each step. A scene illumination spectrum can also be used to convert the radiance into reflectance and remove the influence of varying scene illumination. This spectrum can be derived from measuring the perceived radiance of an object of known reflectance, or through an auxiliary sensor, such as a spectral radiometer, that directly measures the illumination spectrum.

\subsection{Georectification Strategies}

\begin{figure}
\centering
\includegraphics[height=4cm]{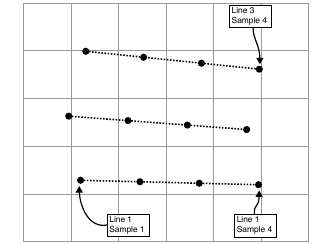}
\caption 
{ \label{fig:mesh_diagram_samp}
Plotting georectified samples from a 4 sample by 3 line cube onto a 6x5 pixel output } 
\end{figure}

Many hyperspectral georectification systems exist in the literature and commercial software, but most are based on a straightforward direct georectification technique~\cite{warren_data_2014}. This method processes each sample of each line independently. First, the ray representing the point in the scene captured to give the particular sample is calculated using the sample number and the sensor parameters. This ray is then transformed by the line's position and orientation to the camera's actual position in the scene (Fig.~\ref{fig:geo_diagram}). The ray is intersected with the ground; this intersection point is the georectified position of the sample and shows where its value came from. Finally, the corresponding value is plotted at the calculated location in the final cube (Fig.~\ref{fig:mesh_diagram_samp}). By repeating this for every sample on every line, the complete georectified cube is built.

This method is conceptually convenient and straightforward to implement, and there are many techniques which build upon it. They commonly create point clouds from the intersections~\cite{inamdar_implementation_2021,finn_automated_2023}, or perform feature matching of larger regions with other image data to improve overall consistency~\cite{angel_automated_2019,yi_seamless_2021}. Nevertheless, direct georectification suffers from fundamental speed and coverage limitations that we address with Spacecube.

While the calculations performed for each sample are not particularly intense (depending on intersection calculation technique), the number of samples is large, and users in the literature generally report times in hours for georectification of datasets, not counting the additional time of their particular technique. There are also some systems which georectify parts of datasets independently, therefore requiring additional time to join together results into a complete cube.

Direct georectification also suffers from incomplete coverage, particularly when aiming to generate results close to the resolution of the input data. Due to the motion of the aerial system, multiple samples may land on the same pixel (wasting information), or a particular pixel may never have a sample land on it (causing a gap in the output). There are interpolation techniques to mitigate gaps, but they require yet more processing time and apply to the cube after georectification, creating inconsistencies and raising questions about interpolated versus ``original" pixels.

\begin{figure}
\centering
\includegraphics[height=4cm]{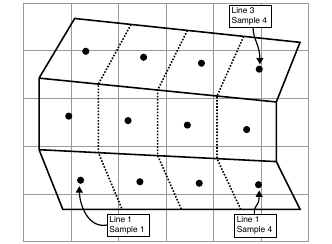}
\caption 
{ \label{fig:mesh_diagram_mesh}
Spacecube expands each sample to the area imaged to collect it, forming a coverage mesh } 
\end{figure}

Spacecube solves the coverage problem by computing a mesh representation of the covered area instead of treating samples as a set of points (Fig.~\ref{fig:mesh_diagram_mesh}). Spacecube then solves the speed problem by rendering this mesh using OpenGL and modern graphics hardware, in essence implementing the inverse of direct georectification. We call our technique the inverse as OpenGL computes the corresponding sample for each georectified pixel, rather than computing the corresponding pixel for each sample to georectify. Provided sufficient storage bandwidth, complete datasets can be georectified in seconds instead of hours. Subsets of the data can be viewed instantly for quality control or investigative purposes.

\subsection{Graphics}
Spacecube leverages the power of modern graphics processing units (GPUs) by using OpenGL to render the generated mesh which represents the georectified data. In OpenGL, meshes are defined by a set of vertices and a list of triangles which connect them.

Each mesh rendering operation renders the component triangles onto a 2D image buffer. The operation can include user-set parameters and other images as input. Rendering data (including vertices, parameters, and images) is processed by shaders, which are small programs written in the C-like GL Shading Language (GLSL)~\cite{rost_opengl_2012}. Though vertices usually describe positions and images usually contain RGB colors, their actual data and its meaning is user-defined and extensible.

For each vertex, the vertex shader calculates its location on the image buffer from its data. For each triangle, the fragment (i.e. pixel) shader then uses data linearly interpolated from the data of its three vertices (plus image data) to calculate the value of each pixel covered by it; these values are stored in the image buffer at the corresponding pixel locations. Since shaders are complete programs processing user-defined data, almost any calculation can be implemented.

Spacecube includes a set of vertex/parameter/image data descriptions and shader programs to implement its rendering, described in Sec.~\ref{subsec:renderer}. Many applications like machine learning and physics simulation also use GPUs, but use different programming APIs like OpenCL and CUDA. As Spacecube is an interactive graphical application that needs rasterization and real-time rendering, we felt comfortable utilizing OpenGL and our tasks are well-solved by its capabilities.

\section{Implementation\label{sec:implementation}}

\begin{figure}
\centering
\includegraphics[width=3.49in]{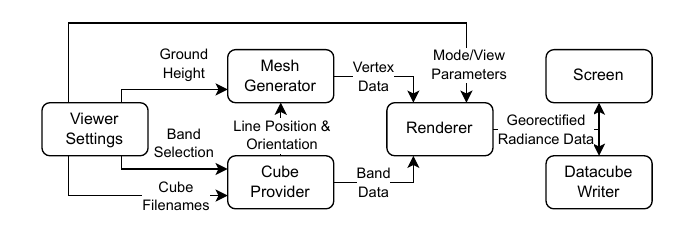} 
\caption 
{ \label{fig:system_overview}
Spacecube component overview } 
\end{figure}

Figure~\ref{fig:system_overview} shows Spacecube's components which together implement a complete georectification solution using our inverse georectification technique. Spacecube is implemented in Python 3.10 and primarily leverages ModernGL 5.8~\cite{dombi_moderngl_2020} to use OpenGL functions. Spacecube's viewer and command line rasterizer are implemented as two separate programs, though most functionality is actually implemented in libraries shared by both and usable in other applications. We focus primarily on describing the viewer, though we note where the rasterizer is different.

Spacecube can be set up to support any line-scan hyperspectral system. It reads and writes datacubes in the industry standard ENVI (ENironment for Visualizing Images) format. Spacecube operates band-sequentially, so only a handful of bands are loaded into memory at a time. Spacecube also operates per-cube; our airborne system by default accumulates 1000 lines into one cube, and each cube is treated as an independent object despite being a part of the same data collection. This facilitates parallel loading and processing of the data and reduces the amount that must be managed at once, improving performance and simplifying implementation.

\subsection{Viewer Operation}

\begin{figure}
\centering
\includegraphics[width=3.49in]{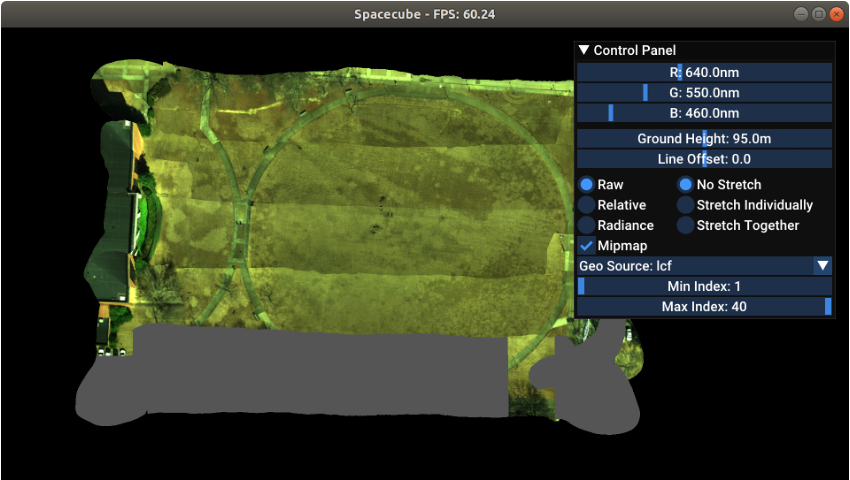}
\caption 
{ \label{fig:viewer}
Screenshot of Spacecube's viewer and controls. Cubes whose data is not yet loaded appear gray }
\end{figure}

While opening the viewer, the user provides a list of cubes to view. Spacecube then loads relevant metadata, processes the georectification information, automatically estimates the ground height, then loads and displays three selected bands on the screen's red, green, and blue channels. The user can zoom and pan around the scene using mouse or touchscreen input, even while loading is in progress. Figure~\ref{fig:viewer} shows the viewer and controls; some cubes are partially loaded for illustration purposes.

The user can set the processing mode to display raw digital number data, relative-scaled data (i.e. scaled to a consistent exposure), or radiance calibrated data (if an appropriate calibration file has been supplied). Each processing mode can be displayed unscaled (most useful for raw data), or scaled (termed as ``stretched"~\cite{driggers_signal_2008}) using one value for all displayed channels or an independent value for each. The scaling values are calculated using a histogram so that pixel values at 2\% or below render as black, and 98\% or above render as white. In raw modes, this histogram is calculated from the raw data; otherwise it is dynamically calculated from the screen content.

The user can select which wavelength to display in each RGB channel. Spacecube looks up the band number whose wavelength is closest to the one requested then loads that band's data. There is a selection for which cubes are displayed on the screen, useful for cropping out starting and ending parts of the flight. The ground height and georectification data source can be changed interactively. The user can optionally enable mipmapping~\cite{williams_pyramidal_1983}, which (slightly) improves performance and reduces aliasing to smooth the image, though it creates small artifacts between cubes.

The viewer is designed to remain responsive for interactive usage. The zoom and pan, histogram, processing mode, and range of displayed cubes update at 60 frames per second. Other operations are processed on background threads and results are displayed to the user as soon as they are available. If the user changes a parameter, an effort is made to cancel existing requests so that the latest data is displayed as quickly as possible. The viewer manages transferring selected band and georectification data to the GPU as it's loaded, limiting the rate to preserve interactivity.

The command line rasterizer always produces unscaled radiance data, never uses mipmapping, and stores results to a band-sequential datacube file rather than displaying them on the screen. It can also process more than three wavelengths; all specified wavelengths get processed in sequence.

\subsection{Provider}

Spacecube abstracts retrieving cube data to process through a cube provider library, so-named as it provides the data Spacecube processes. This allows different sources to be used in the future, for different data formats and different types of experiments. Spacecube supplies filenames (or other sources) of cubes to the provider, then the provider handles finding the cubes and converting the data into a common format for Spacecube to use.

The provider initially loads cube metadata only (such as dimensions and camera parameters), then provides a handle that Spacecube uses to retrieve a particular cube band or set of georectification data when necessary.

The provider, like Spacecube, is designed for band-sequential cubes; requests are processed one band at a time. Other orderings are supported, but performance can be poor due to problematic disk access patterns (when reading a particular band, accessing the first 1/300th of a file's bytes is dramatically faster than accessing every 300th byte). There is functionality to pre-load all data into memory to mitigate these problems on sufficiently capacious computers.

The provider also handles pre-processing of georectification data. The data is read from the INS files and initially formatted into a capture timestamp, the camera's geographic position (latitude, longitude, altitude), and the camera's orientation (roll, pitch, yaw).

Latitude and longitude aren't suitable to operate on directly, as they represent a spherical coordinate system, but we want to render nominally flat maps. Therefore, they are projected to the Universal Transverse Mercator~\cite{snyder_map_1987} (UTM) coordinate system. The appropriate UTM zone and grid heading offset are automatically picked when data is loaded; we assume sUAS do not have flight areas big enough that distortion due to a non-ideal zone or offset becomes a problem. After projection, the coordinates are stored internally as meters in a North-East-Down (NED) coordinate system.

The orientation data is transformed so that the identity rotation represents the camera aiming nadir with the top of the capture line pointing North in the NED system. The calculated UTM heading offset is applied, then the data is stored internally as a SciPy 1.11 Rotation object (holding multiple rotations).

Linear interpolation of position data and spherical linear interpolation of orientation data is then performed using the exposure and INS data timestamps so that each line of the cube is assigned the position and orientation of the exact start of its exposure. This data is then given to Spacecube as an array of NED coordinates in meters and a Rotation object, each containing one entry per line in the cube. This pre-processing ensures Spacecube can operate using a consistent and known coordinate system in both space and time, independent of the cube or georectification data source.

\subsection{Mesh Generator}

Before rendering each cube, Spacecube generates the mesh that represents the ground area the camera imaged while collecting the cube. The mesh does not contain the actual cube data and is generated independently of it. For this generation, we assume the camera flies over a flat surface that is at a known altitude (i.e. ground height), which is a reasonable assumption for areas like crop fields scanned by sUAS. We also assume the camera is exposing continuously so that there are no gaps between lines, which is an operation mode supported by most cameras.

Note that our meshing process treats a particular sample as covering a certain area, rather than being placed at a particular point as in the direct georectification method. We believe this is more accurate because (we assume) the camera exposes continuously during the line, and is therefore collecting information from the entire area as it moves between lines. The area covered by a particular sample in the line is also not infinitesimally small and is instead described by the physical detector size and optics; again we assume there is no gap or overlap. However, we still compute the coordinates of each line's two endpoints on the ground similarly to direct georectification to identify the bounds of the area covered by the cube.

Starting with the camera in the identity orientation, two vectors are (de)projected from the origin using the pinhole camera model~\cite{szeliski_computer_2022}, one from each extreme of the camera's field of view, so that their heads land on a surface 1 unit below. Knowing the camera field of view angle $\theta$ and using straightforward trigonometry, the horizontal distance from each head to a vector of length 1 that points straight down is $d = 1*\mathrm{tan}^{-1}(\frac{\theta}{2})$. The two vector heads are then $(N, E, D) = (0, \pm d, 1)$.

For each cube line, the orientation and position at the start of its exposure is retrieved. The two vectors are rotated to that orientation and translated to that position so their heads are at the 3D endpoints of the image projected onto the camera sensor when capturing that line. For perspective correct interpolation~\cite{heckbert_interpolation_1991}, we compute an effective depth of each head from the vertical distances of that head to the camera and to the ground. Both vectors are finally extended along their direction to intersect with the ground; these 2D intersection points are the endpoints of the line.

\begin{figure}
\centering
\includegraphics[height=4cm]{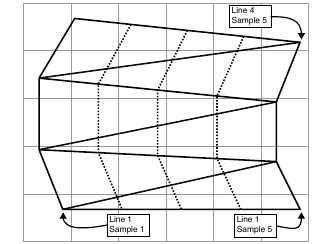}
\caption 
{ \label{fig:mesh_diagram_tris}
The sample coverage area expressed as a triangle mesh with vertices at line endpoints (dotted sample dividers for illustration only)} 
\end{figure}

To cover the area imaged, the two endpoints of a line are joined to each other and to the two endpoints of the next line with a quad whose vertices are the four points (expressed naturally as two triangles). By repeating this in sequence for all lines in the cube, the entire area imaged by the camera is covered by a series of triangles which together form a gap-less mesh (Fig.~\ref{fig:mesh_diagram_tris}). The vertices are then tagged with their depth and uploaded to the GPU for rendering.

To avoid a gap after the cube and account for its full area, the first line of the next cube in the capture sequence is treated as the ``next" line of the current cube's last line. A complete data collection will therefore be rendered as if it were one gap-less mesh despite being composed of many cubes with independently generated meshes. At the end of the collection, where no next cube exists, the last line is assumed to be zero area and thus not rendered.

\subsection{Renderer \label{subsec:renderer}}

The renderer processes three bands at once (mapped to R, G, and B channels) and renders onto a buffer with pixel dimensions given by the viewer window size or rasterization ground sample distance (GSD)~\cite{driggers_encyclopedia_2003}. The render area of the buffer is described as a center (in North-East coordinates) and a scale (in meters) for each dimension. The viewer changes the center and scale in response to user pan and zoom commands; the rasterizer calculates them based on the provided cube bounds.

Each cube mesh is rendered in one operation. The vertex data comes from the line endpoint positions and depths. Parameters are used to describe the viewing area and calibration mode. Images are set up that contain the band data, calibration data, and the per-line camera response calculated from the camera settings.

The vertex data is transformed by the vertex shader to screen coordinates according to the view parameters, avoiding the need to change the data in memory and so improving performance. The vertex shader also adds to each vertex its line number and sample number, calculated from the known mesh structure. 

\begin{figure}
\centering
\includegraphics[height=4cm]{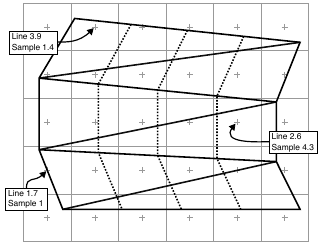}
\caption 
{ \label{fig:mesh_diagram_interp}
OpenGL interpolates vertex data to provide cube location of each covered pixel
(dotted sample dividers for illustration only) } 
\end{figure}

OpenGL computes the pixels covered by the mesh, then interpolates the vertex data to calculate cube positions at the center of these pixels (Fig.~\ref{fig:mesh_diagram_interp}). The cube position is interpolated perspectively-correct, which assumes the quad connecting the line to its next is flat in 3D space, and so renders the lines that divide samples as straight. However, if there is a difference in roll between the two lines, the quad cannot be flat, and it will appear creased along the edge that joins the quad's two triangles. In practice, the position error from this effect averages less than 1\% the size of a sample. By interpolating here instead of after georectification, each pixel value is guaranteed to exist and be geometrically accurate, instead of missing or possibly filled in from ``nearby".

At each covered pixel, the fragment shader uses the calculated line and sample number to look up the appropriate band data value, calibration coefficients, and camera response from the input images. As discussed in Sec.~\ref{sec:calibration}, these values are applied to produce the radiance value according to the formula
\begin{equation}
\label{eq:fov}
(\mathrm{cube}_{band, line, sample} - \mathrm{calib}^{\mathrm{dark}}_{band, sample}) * \frac{\mathrm{calib}^{\mathrm{rad}}_{band, sample}}{\mathrm{response}_{line}} \, .
\end{equation}
Inputs change depending on mode, e.g. disabling calibration fixes $\mathrm{calib}^\mathrm{dark} = 0$ and $\mathrm{calib}^\mathrm{rad} = 1$.

This value is then scaled to the display range, stored to the buffer at the pixel's location, and eventually displayed on the screen or written to a datacube. This process is performed independently for each of the three selected bands.

\section{Results}

We describe the parameters of the aerial line-scan hyperspectral system we used and a dataset we have collected with it. We then process this dataset with Spacecube, which uses our inverse georectification technique, along with other software that uses direct georectification, in order to compare processing performance. Finally, we compare quality of the results of each program.

\subsection{Data Collection}

For collection we used a Resonon Pika L~\cite{noauthor_resonon_2024} airborne hyperspectral system, modified and adapted to fit our sUAS and improve reliability. The system includes an SBG Systems Ellipse N~\cite{noauthor_sbg_2023} INS and an Emlid Reach RS2+~\cite{noauthor_emlid_2024} GNSS base station to provide real-time kinematic (RTK) correction data. The Pika L camera produces lines with 900 samples and 300 bands at a rate of 249 frames per second and exposure time of 3.9ms (97\% of frame interval time). The INS produces location and orientation data at a rate of 200Hz. The system was flown on a custom-built 1000mm-class hexa-rotor sUAS running ArduPilot~\cite{tridgell_ardupilot_2024}.

To generate comparable results, we primarily used Resonon's hyperspectral analysis program, Spectronon v3.5.5, which is designed for and provided with the Pika L camera we used. Spectronon has plugins to perform radiometric calibration, georectification using the direct georectification method, and mosaicing into a final datacube. We also used the industry standard PARGE~\cite{schlapfer_parge_1998,noauthor_parge_2024} v4.1b7 tool, which is partially compatible with the Pika L and also uses the direct georectification method.

Using our collection system, we collected a dataset at the University of Memphis on 2024-02-23 at 11:00 AM. The flight area was roughly 175m by 125m (2.2ha), and was collected over 5 passes at a nominal altitude of 40m (above a ground height of 95m) and flight speed of 10m/s using a 47.5 degree field of view lens. Using these flight parameters, nominal ground sample distance (GSD) was approximately 4cm in both axes. We therefore configured 4cm as GSD and 95m as ground height for all processing.

The total used data comprises 40 cubes, containing 40,000 lines of data, collected over 2 minutes and 40 seconds. Additional data collected during takeoff, in-flight calibration, and return/landing was not processed to avoid distractions and artifacts.

\begin{figure}
\centering
\includegraphics[width=3.49in]{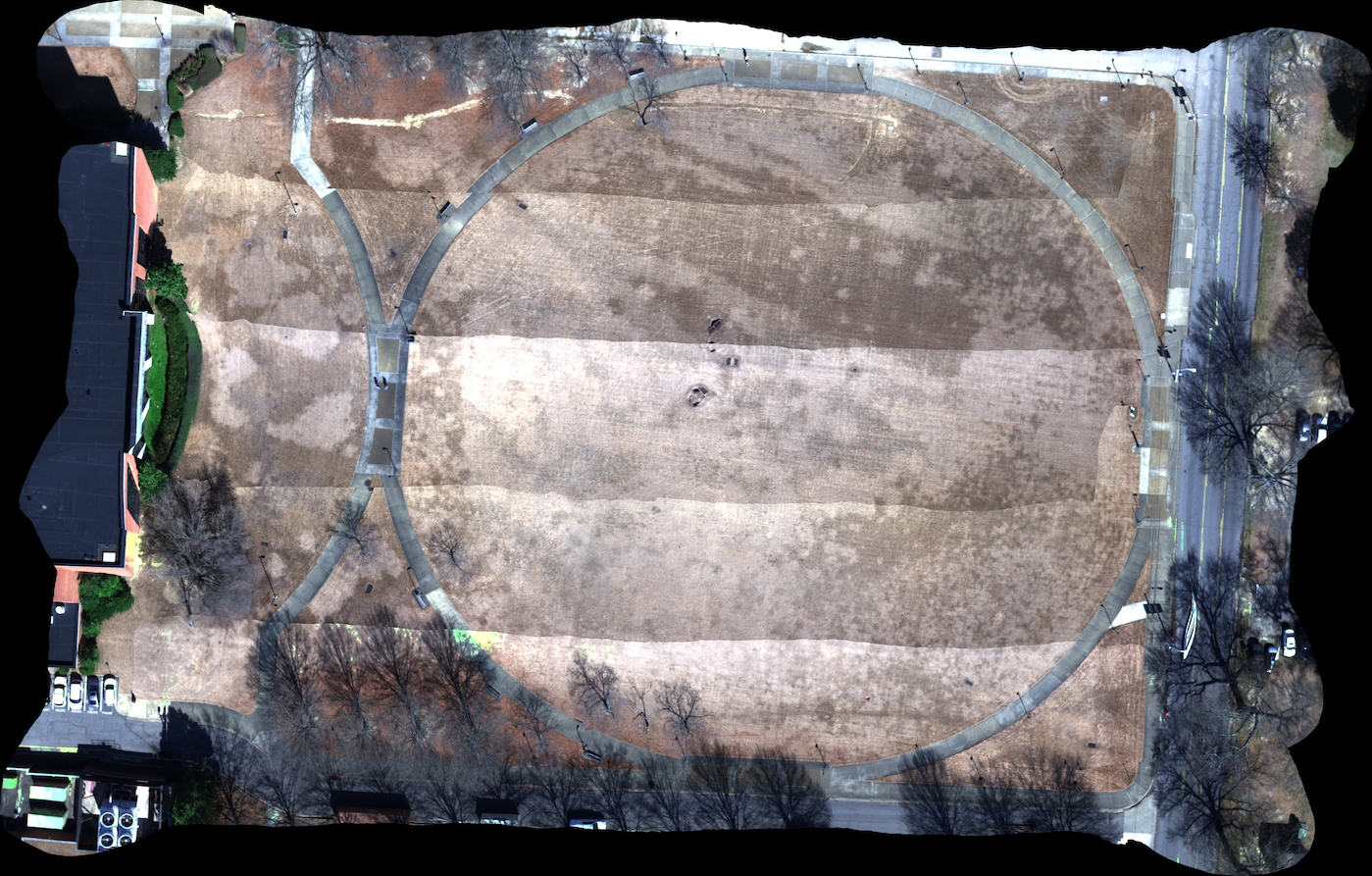}
\caption 
{ \label{fig:data_final}
Georectified result }
\end{figure}

The complete georectified result, as processed by Spacecube, is shown in Fig.~\ref{fig:data_final}. There are some artifacts where the cube rows overlap due to lens vignetting and perspective effects on a not-quite-flat surface. There are also some small green splotches, particularly in the lower left corner, due to over-exposure. This result is, by design, essentially indistinguishable to that produced by Spectronon, so we don't spend space showing Spectronon's output. However, we will focus in on certain sections of this result to highlight where our approach excels. 

\subsection{Performance Comparison}

We measured processing time of each step of each program on a ThinkPad P52 laptop with an Intel Xeon E-2176M CPU (6 cores, 2.7GHz, 64GB RAM) and an Nvidia Quadro P2000 GPU (4GB VRAM). To avoid disk access patterns and caches influencing results, we keep all data files in a RAM disk for all tested software. The results are summarized in Table~\ref{tab:results}.

Spacecube's rasterizer command line program generated the complete 300 band cube in 11 seconds, including application of the radiance calibration, georectification, and mosaicing into a final datacube. This is 1,400\% the speed of real time operation, as the data was collected over 2 minutes and 40 seconds.

Spectronon required 7 minutes and 16 seconds to apply the radiance calibration and perform georectification of each of the 40 cubes individually, plus an additional 14 minutes and 13 seconds for mosaicing. Spacecube is over 11,000\% the speed of Spectronon and can perform all the steps at once without any intermediate files.

PARGE does not implement calibration for the Pika L, so Spectronon's calibration result was reused. PARGE needed 3 minutes and 11 seconds to georectify all the cubes (using its fast flat ground assumption); this time was estimated by processing a subset of the cubes then scaling the time to the total cube count, as PARGE's batch processing doesn't support the Pika L. PARGE then needed nearly 30 minutes for mosaicing (again re-using Spectronon's georectification result, with format conversion). Spacecube is over 19,000\% as fast as PARGE in PARGE's fastest mode.

\begin{table}[!t]
\caption{Processing Time (minutes and seconds)} 
\label{tab:results}
\small
\centering
\begin{tabular}{|l|c|c|c|}
\hline
\rule[-1ex]{0pt}{3.5ex}  Processing Step & Spectronon & PARGE & Spacecube  \\
\hline\hline
\rule[-1ex]{0pt}{3.5ex}  Radiance Correction & \multicolumn{2}{c|}{4m 15s} & \multirow[c]{3}{*}{0m 11s}  \\
\cline{1-3}
\rule[-1ex]{0pt}{3.5ex}  Georectification & 3m 1s & 3m 11s (est.) & {}   \\
\cline{1-3}
\rule[-1ex]{0pt}{3.5ex}  Mosaicing & 14m 13s & 28m 22s & {}   \\
\hline\hline
\rule[-1ex]{0pt}{3.5ex}  Total & 21m 29s & 35m 48s & 0m 11s  \\
\hline 
\end{tabular}
\end{table} 

\subsection{Quality Comparison}

\begin{figure}
\centering
\includegraphics[width=3.49in]{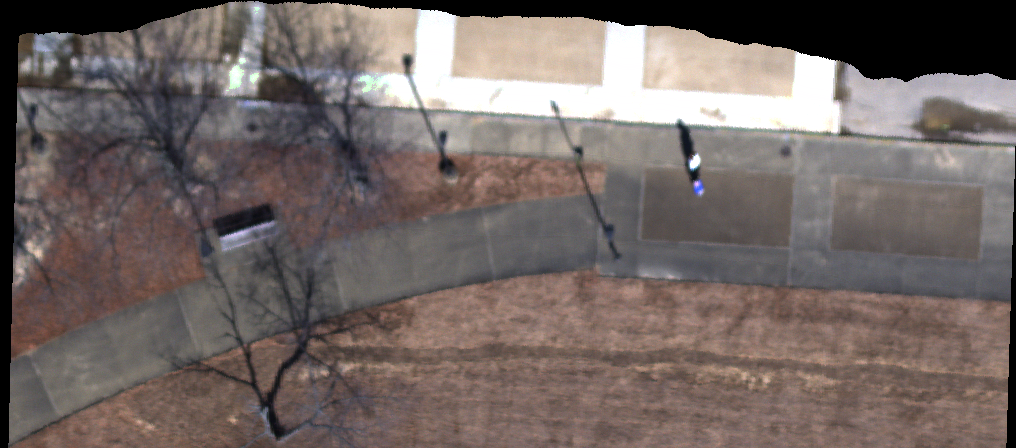}
\caption 
{ \label{fig:cube_26_spectronon_interp}
Spectronon's georectification can produce visually indistinguishable results } 
\end{figure}

\begin{figure}
\centering
\includegraphics[width=3.49in]{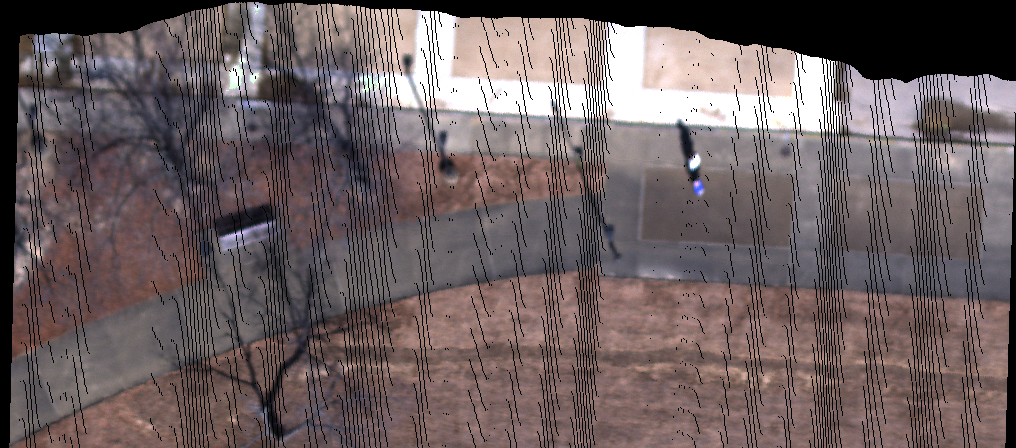}
\caption 
{ \label{fig:cube_26_spectronon_nointerp}
Disabling Spectronon's interpolation reveals gaps, resulting in striped coverage artifacts } 
\end{figure}

Using Spectronon's default interpolation settings (morphological with 1 pixel radius), we georectify dataset cube 26 (shown in the introduction) and display the result in Fig.~\ref{fig:cube_26_spectronon_interp}. This cube is visually indistinguishable from one produced by Spacecube in Fig.~\ref{fig:geo_after}. However, turning off interpolation (Fig.~\ref{fig:cube_26_spectronon_nointerp}) results in gaps that create striped patterns of uncovered pixels. In both cases, PARGE produces results visually indistinguishable from Spectronon's as it also uses direct georectification.

These gaps in the georectification are pixels not covered because no sample happened to land there due to small variations in the sUAS orientation. Gaps are intrinsic to the direct georectification method, and are especially likely when the output GSD is similar to the nominal input GSD. Interpolation then covers gaps afterwards by simply replacing each with the value of a nearby pixel. For our 4cm GSD dataset, Spectronon suggests a GSD of 7cm, which does reduce gaps. However, the 75\% increase in pixel edge length of course enlarges their area, causing a 300\% resolution reduction.

\begin{figure}
\centering
\includegraphics[width=3.49in]{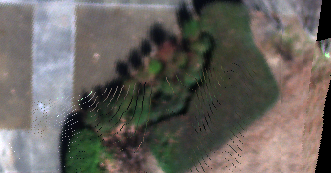}
\caption 
{ \label{fig:cube_23_spectronon_interp}
Patterns of incorrect pixels are visible despite Spectronon's interpolation } 
\end{figure}

\begin{figure}
\centering
\includegraphics[width=3.49in]{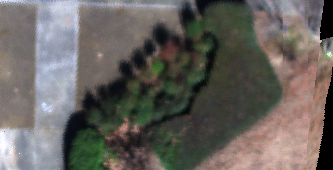}
\caption 
{ \label{fig:cube_23_parge_interp}
PARGE has fewer gaps and incorrect pixels, though some are still present in the lower left } 
\end{figure}

While interpolation does mitigate gaps, it has weaknesses, particularly with high motion and in overlapping areas. Figure~\ref{fig:cube_23_spectronon_interp}, showing dataset cube 23 processed by Spectronon, has radial striping in the bottom half where two orientations overlap as the sUAS makes a turn. The interpolator does not see these as gaps because a sample has landed on them. Since it runs after georectification, the interpolator cannot recognize that the sample is inconsistent from the perspective of the whole image. PARGE surprisingly generates fewer incorrect pixels at the overlap, though there are still some in the lower left (Fig.~\ref{fig:cube_23_parge_interp}).

\begin{figure}
\centering
\includegraphics[width=3.49in]{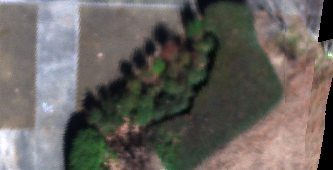}
\caption 
{ \label{fig:cube_23_spacecube}
Spacecube's mesh rendering naturally avoids coverage artifacts } 
\end{figure}

Spacecube's rendering method (Fig.~\ref{fig:cube_23_spacecube}) inherently does not produce gaps, eliminating the need for an interpolation pass and completely avoiding any coverage artifacts. The generated mesh fully covers all captured samples, and OpenGL guarantees that all pixels covered by the mesh will be filled in a geometrically accurate manner even if no particular sample lands on a given pixel.

\section{Conclusion}

Spacecube's inverse georectification technique, in contrast with the typical direct georectification method, enables over 11,000\% faster data processing and is immune to artifacts due to pixel coverage gaps. The interactive viewer makes exploratory data analysis easy through fast panning and zooming, in combination with live tweaking of aspects like georectification data and radiance processing mode. The command line tool rapidly generates datacubes in the industry standard ENVI format with the same quality as the viewer. Even when processing all cubes and bands of a dataset, Spacecube operates 1,400\% faster than real time, facilitating new use cases.

Spacecube's biggest limitation is its assumption that the ground is flat. While reasonably valid for the crop surveys we do with sUAS, there are artifacts where passes overlap due to the ground not being perfectly flat, combined with perspective effects. Direct georectification software, including Spectronon and PARGE, can easily use a digital elevation map (DEM) to compensate for these effects. Spacecube cannot, and augmenting the algorithm to use one is non-trivial. Nevertheless, this is a big area of future work for us. Spacecube also does not yet implement averaging of overlapping samples, but this is supported by OpenGL through blending modes, and we plan to add this functionality as well.

\appendix 

\subsection*{Disclosures}
The authors declare that there are no financial interests, commercial affiliations, or other potential conflicts of interest that could have influenced the objectivity of this research or the writing of this paper.

\subsection* {Code, Data, and Materials Availability}
The source code for the version of Spacecube with the capabilities demonstrated in this paper will be made available under a GPLv3 (or later) license at \url{https://github.com/JacobsSensorLab/spacecube-paper-release} after peer-reviewed publication. Spectronon is freely available for download from Resonon's website, though the georectification plugin is not. A free trial of PARGE is available by contacting the developer, ReSe Applications GmbH. The dataset used is available at Zenodo under a CC-BY-4.0 license at \url{https://zenodo.org/records/14814175}.

\section* {Acknowledgments}
We would like to specially thank the lab members who assisted with the design and operation of the drone and airborne system used to collect our data, including Angelin Favorito, Kevin McKenzie, Brianna Miller, Joe Perry, MD Rafi Ur Rahman, Shoaf Robinson, Noah Wargo, and Ryan Williamson. Without their help, we would have had nothing to georectify.

We would also like to thank Peter Le, Patrick Leslie, and Kevin McKenzie for their review of earlier drafts and valuable writing input and encouragement.

This research was sponsored by the Army Research Laboratory and was accomplished under Cooperative Agreement Number W911NF-21-2-0294. The views and conclusions contained in this document are those of the authors and should not be interpreted as representing the official policies, either expressed or implied, of the Army Research Office or the U.S. Government. The U.S. Government is authorized to reproduce and distribute reprints for Government purposes notwithstanding any copyright notation herein.


\bibliographystyle{IEEEtran}
\bibliography{IEEEabrv,report}


\begin{IEEEbiography}[{\includegraphics[width=1in,height=1.25in,clip,keepaspectratio]{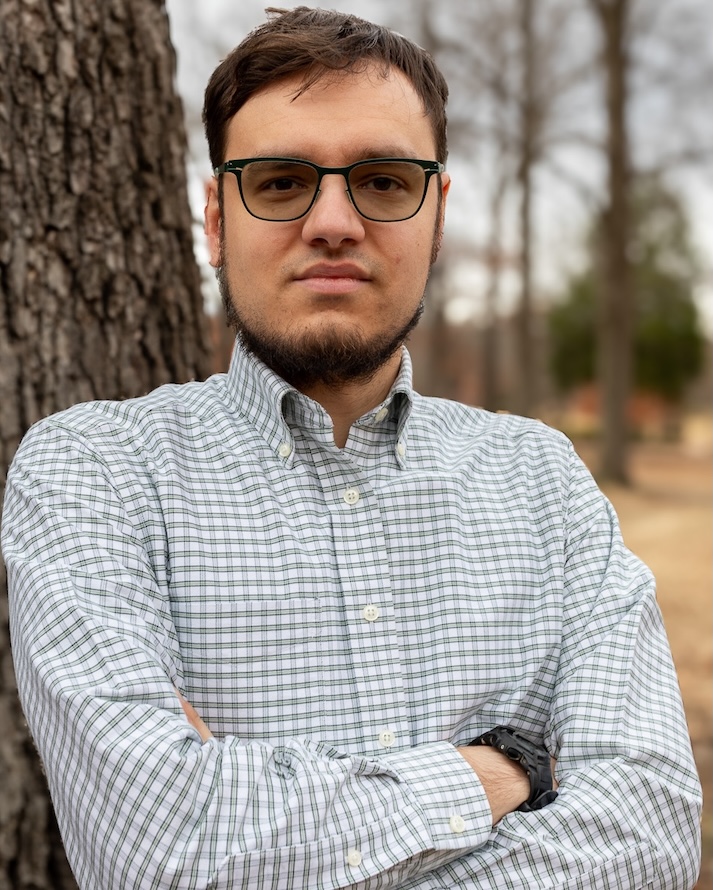}}]{Thomas P. Watson} received his BSEE+CE, and MSEE from the University of Memphis in 2019 and 2021 respectively, and is currently working on his PhD. He likes programming, embedded systems, and system integration, particularly combining old stuff in new ways. His research interest is in flight systems and remote sensing, particularly incorporating real-time processing.
\end{IEEEbiography}

\begin{IEEEbiography}[{\includegraphics[width=1in,height=1.25in,clip,keepaspectratio]{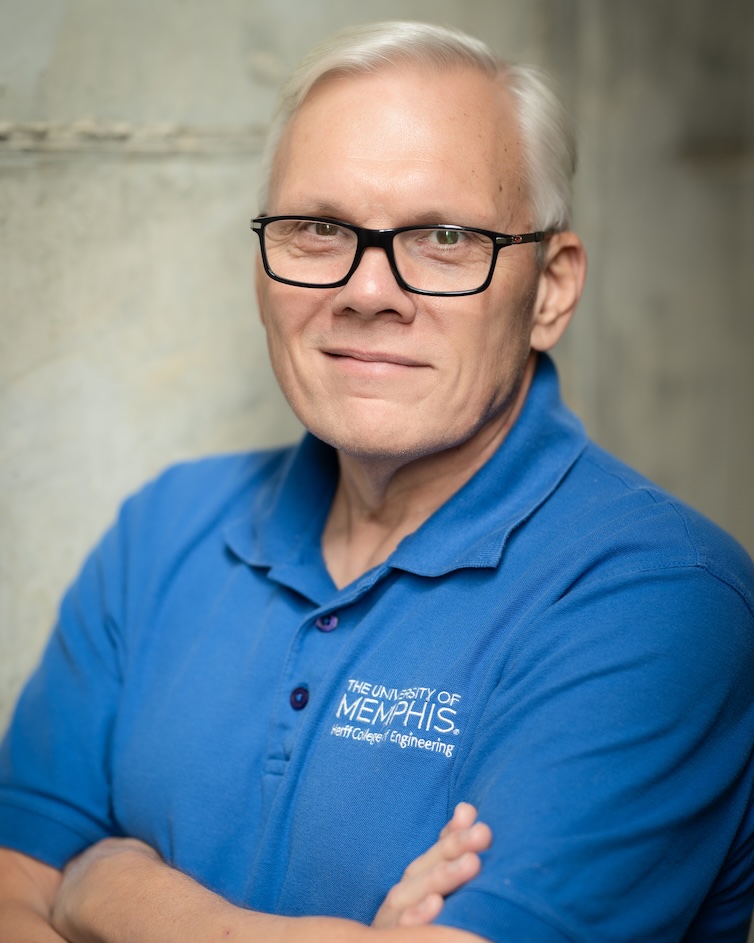}}]{Eddie L. Jacobs}
received the BSEE and MSEE from the University of Arkansas in 1986 and 1988 respectively. He received the Doctor of Science in Electro-physics from The George Washington University in 2001. From 1989 to 2006 he was employed at the US Army Night Vision and Electronic Sensors Directorate, Ft. Belvoir VA. His duties included electro-optic modeling and simulation of imaging sensors. Since 2006 he has served as a professor of Electrical and Computer Engineering at the University of Memphis. His research interests are electro-optic and imaging sensors and remote sensing.
\end{IEEEbiography}

\end{document}